\documentclass[referee]{raa}            

\usepackage{booktabs,threeparttable}
\usepackage{multirow}
\usepackage{longtable}
\usepackage{graphicx,times}             
\input{epsf.sty}                        
\input{psfig.sty}                       

\begin{document}

   \title{X-ray and Optical Plateau Following the Main Bursts in Gamma-Ray Bursts and SNe II-P: A hint to the similar late injection behavior?
}

   \volnopage{Vol.0 (200x) No.0, 000--000}      
   \setcounter{page}{1}          

   \author{X.-H. Cui
      \inst{1}
      \and R.-X. Xu
      \inst{2}
   }

   \institute{National Astronomical Observatories, Chinese Academy of Sciences,
             Beijing 100012, China; {\it xhcui@bao.ac.cn}\\
                \and
             School of Physics and State Key Laboratory of Nuclear Physics
and Technology, Peking University, Beijing 100871,
China\\
   }

   \date{Received~~2009 month day; accepted~~2009~~month day}

\abstract{We analyze the emission plateaus in the X-ray afterglow lightcurves of gamma-ray bursts (GRBs) and in the optical lightcurves of Type II superpernovae (SNe IIP) in order to study whether they have similar late energy injection behaviors. We show that correlations of bolometric energies (or luminosities) between the prompt explosions and the plateaus for the two phenomena are similar. The Type II SNe are in the low energy end of the GRBs. The bolometric energies (or luminosities) in prompt phase $E_{\rm expl}$ (or $L_{\rm expl}$) and in plateau phase $E_{\rm plateau}$ (or $L_{\rm plateau}$) share relations of $E_{\rm expl} \propto E_{\rm plateau}^{0.73\pm 0.14}$ and $L_{\rm expl} \propto L_{\rm plateau}^{\sim 0.70}$. These results may indicate a similar late
energy injection behavior to reproduce the observed plateaus of the plateaus in the two phenomena.
 }

   \authorrunning{X.-H. Cui \& R.-X. Xu }            
   \titlerunning{Plateaus of gamma-ray bursts and supernovae: A unified correlation?}  

   \maketitle

%
%
\section{Introduction}           
\label{sect:intro}

One of the big problems for today's astrophysicists is to understand the explosive mechanisms of Gamma-ray bursts (GRBs) and core-collapse supernovae. Very interestingly, a radiation plateau in the X-ray/optical bands after GRBs and an optical plateau after initial bursts of SNe IIp are usually detected. We focus on this feature
and study the possible relations of the plateaus with the initial bursts for the two kind of events.

On one hand, the early X-ray afterglow of GRB is found to show a
canonical behavior (Zhang et al. 2006; Nousek et al. 2006) by X-Ray
Telescope (XRT) on {\em Swift}. As one of the components in this
canonical X-ray light curve, the shallow decay phase, i.e.
``plateau'', typically lasts a few thousands of seconds with a
temporal decay slope $\sim$-0.5. Various kinds of models, such as
the energy injection model (Rees \& M{\'e}sz{\'a}ros 1998; Nousek et
al. 2006; Zhang et al. 2006), the reverse shock model (Genet et al.
2007), two component model (de Pasquale et al. 2009), the dust
scattering model (Shao \& Dai 2007) etc, have been proposed to
explain this mystic phase. However, a chromatic behavior, i.e., no optical break or
spectral evolution at the transition time ($t_{\rm tr}$) from the
plateau to the normal decay phase in more than half bursts (Fan \&
Piran 2006; Liang et al. 2007), is very difficult to interpret
within the framework of the external shock models (Fan \& Piran
2006; Panaitescu et al. 2006). A suppressed forward shock emission
is required for long lasting reverse shock models (Genet et al.
2007; Uhm \& Beloborodov 2007). The spectral evolution could not be
interpreted by dust scattering effect (Shao \& Dai 2005) though the
lightcurve can be explained. Also the two-component external shock
jets (de Pasquale et al. 2009) would require contrived shock
parameters. A long-lasting central engine therefore possibly
explains the X-ray plateau phase in GRB afterglow emission and is
concerned by the chromatic scenario (Liang et al. 2007). From the
observations, the isotropic X-ray energy ($E_{\rm iso, X}$) for the
plateau phase in afterglow of GRB is found to be correlated
with the prompt gamma-ray energy and the transition time $t_{\rm
tr}$ (Liang et al. 2007). An anti-correlation has been found between
the end time of the plateau $T_{\rm a}$ and the X-ray luminosity
($L_{\rm X}$) at $T_{\rm a}$ in the GRB rest frame (Dainotti et al.
2010). By adding a third parameter, i.e. the isotropic $\gamma$-ray
energy $E_{\rm{iso}}$, Xu \& Huang (2011) found a new and
significantly tighter three-parameter correlation for gamma-ray
bursts with a plateau phase in the afterglow.

On the other hand, plateaus also appear in the light curve of type
II Plateau supernovae (SNe II-P). Observationally, SNe II-P are
classified as a ``plateau'' on the slow decay of their early light
curves (Barbon et al. 1979), where the luminosity remains nearly
constant for a period of $\sim$70-100 days (Pskovskii 1978). Their
expansion velocities, plateau luminosities and durations show a wide
range (Young \& Branch 1989; Hamuy 2001). In order to reproduce the
plateaus of SNe II-P, a red supergiant progenitor with an extensive
H envelope would be necessary (Grassberg et al. 1971; Falk \& Arnett
1977). An analytic model (Arnett 1980; Popov 1993) and hydrodynamic
models (Litvinova \& Nadezhin 1983, 1985) are introduced to explain
the light curves of SNe II-P and their correlation with the physical
parameters of progenitor stars. It is conventionally accepted that
the plateau phase in type II-P supernovae results from the
recombination of ionized hydrogen. However, the way of the diffusion
photons through the expanding envelope after the shock reaches the
surface and the mechanism of energy deposition in the envelope still
remains unknown though many efforts have been tried to study the
structure and the hydrodynamic of envelope after the core collapse
of center star (Arnett 1980; Popov 1993; Litvinova \& Nadezhin 1983,
1985; Burrows et al. 2006; Janka et al. 2007; Utrobin \& Chugai
2009).

It is known that some long GRBs are associated with core-collapse supernovae
(SNe). The discovery of 30 associations between long, soft
GRBs and Type Ib/c SNe (see, e.g., the review by Woosley \&
Bloom 2006 and Hjorth \& Bloom 2012) directly tell that their progenitors are massive stars. And these associations result in finding common explosive processes for
SNe and GRBs to form rapidly spinning black holes (Woosley
1993), neutron stars (Klu¡äzniak \& Ruderman 1998), or even
quark stars (Dai \& Lu 1998a). And a quantitative relation between the peak spectral energy of GRB and the peak bolometric luminosity of
the SN was also presented to clarify that the critical parameter determining the GRB-SN connection is
the peak luminosity of SNe (Li 2006). In the standard collapsar model of GRBs, collimation of the outflow is essential for avoiding baryon loading and
producing a clean fireball. However, for some GRBs/XRFs, the jet opening angle inferred from the correlation between the jet opening angle of GRBs and the peak energy of their spectra measured in the GRB frame is so large that the burst outflow should be spherical (Li 2006). This is consistent with radio
observations on the soft XRF 020903, GRB 060218, and XRF 080109 (Soderberg et al. 2004, 2006, 2008). Two possible scenarios for producing a GRB/XRF from a spherical configuration have been presented (Li 2008).

Comparative studies of plateaus in GRB afterglows and those in SNe
II-P can reveal their properties, hydrodynamics and the possible
physical process/origins. This work is to show implication about if
there is a correlation for the plateau phenomena and if
there is a similar hydrodynamical process or energy injection behavior during the plateau phase. In
this paper, we analyze the observed parameters for 43 {\em Swift}
XRT GRB afterglows and those for 11 SNe II-P collected from prior
work. A correlation between the energies $E_{\rm expl}$ in prompt phase and $E_{\rm{plateau}}$ (i.e. $E_{\rm{plateau}} \times \tau$, where $\tau$ is the duration of plateau phase) in the plateau phase has been found for both sample. The relation between luminosity $L_{\rm expl}$ (i.e. $E_{\rm expl}/\tau$) and the $L_{\rm{plateau}}$ can also be well fitted with a power law. The power-law indies of both correlations are found to be similar for two samples within the error bar ranges.
This may imply a similarity between the dynamic processes or energy injection behavior to
reproduce the plateaus during those two kinds of explosions of GRB
afterglow and SNe II-P though in different regime. The energy
budgets for plateau and (prompt) explosion are correlated for both
samples, respectively. The data of samples and the calculation method are
presented in \S 2. The bolometric luminosity is deduced from the
fitting of the lightcurve for GRB X-ray afterglow and SNe II-P. In
order to compare the properties of plateau, in \S 3 we present two
correlations between the luminosities $L_{\rm
expl}$ and $L_{\rm plateau}$, as well as the energies $E_{\rm
expl}$ and $E_{\rm plateau}$, for GRB and
SNe II-P samples. The results are summarized in \S4 with some
discussion.

\section{Data and Method}
\label{sect:Data and Method}

The X-ray afterglow of our GRB sample is downloaded from the {\em Swift}
XRT data archive. The redshift of the bursts in this GRB sample are all detected up to 2010 December. And the sample includes only those XRT light curves with a clear initial steep decay segment, a shallow
decay segment and a normal decay segment detected by {\em Swift}/XRT.
We get a sample of 43 GRBs including the 33 GRBs in the work of Cui et al. (2010) and
another 10 bursts after November of 2008 as shown in Table
\ref{tabGRB}.
\begin{longtable}{rcccccc}
\caption{The properties of GRB sample\label{tabGRB}}\\
  \toprule & \\
\midrule
\endfirsthead 
\midrule & \\
\midrule
\endhead 
\midrule
{continue goes here\dots} \\
\endfoot 
\endlastfoot
 GRB & $z$ & $T_{90}$ & $\Gamma_X$ &$\tau_{\rm GRB}$ & $L_{\rm plateau,GRB}$ &$E_{\rm expl,GRB}$ \\
& &(s) & &(ks)& ($\times 10^{48}$erg/s)&($\times 10^{53}$ erg)  \\
 \hline
050416A	&	0.65	&	2.4	&	2.15	&	0.17	$\pm$	0.11	&	0.07	$\pm$	0.05	&	0.01	\\
050803	&	0.42	&	110	&	1.88	&	1.24	$\pm$	0.09	&	0.03	$\pm$	0.02	&	0.03	\\
050908	&	3.35	&	19.4	&	3.9	&	0.52	$\pm$	0.14	&	0.27	$\pm$	0.87	&	0.35	\\
051016B	&	0.94	&	4	&	2.82	&	6.85	$\pm$	2.31	&	0.02	$\pm$	0.01	&	0.01	\\
051109A	&	2.346	&	14.3	&	2.33&	0.58	$\pm$	0.14	&	3.59	$\pm$	1.50	&	0.90	\\
060108	&	2.03	&	14.4	&	1.91&	2.23	$\pm$	0.74	&	0.12	$\pm$	0.10	&	0.12	\\
060210	&	3.91	&	255	&	1.93	&	0.59	$\pm$	0.15	&	15.70	$\pm$	5.08	&	6.91	\\
060418	&	1.49	&	103.1	&	2.04&	0.06	$\pm$	0.02	&	3.03	$\pm$	1.28	&	1.57	\\
060502A	&	1.51	&	33	&	2.43	&	5.12	$\pm$	1.51	&	0.18	$\pm$	0.06	&	0.45	\\
060510B	&	4.9	&	275.2	&	1.42	&	13.41	$\pm$	3.25	&	0.07	$\pm$	0.19	&	5.05	\\
060522	&	5.11	&	71.1	&	1.97&	0.05	$\pm$	0.02	&	6.13	$\pm$	31.03	&	1.50	\\
060526	&	3.21	&	298.2	&	1.8	&	1.11	$\pm$	0.28	&	0.53	$\pm$	0.56	&	0.85	\\
060605	&	3.8	&	79.1	&	1.6	&	    0.58	$\pm$	0.15	&	2.06	$\pm$	3.17	&	0.6	\\
060607A	&	3.08	&	100	&	1.79	&	1.20	$\pm$	0.02	&	11.61	$\pm$	3.33	&	1.61	\\
060707	&	3.43	&	66.2	&	2	&	0.64	$\pm$	0.16	&	1.08	$\pm$	1.34	&	1.19	\\
060708	&	2.3	&	9.8	&	2.51	&	    0.59	$\pm$	0.38	&	0.61	$\pm$	0.40	&	0.20	\\
060714	&	2.71	&	15	&	2.02	&	0.53	$\pm$	0.10	&	1.49	$\pm$	1.28	&	1.47	\\
060729	&	0.54	&	116	&	2.71	&	6.95	$\pm$	0.30	&	0.03	$\pm$	0.01	&	0.07	\\
060814	&	0.84	&	146	&	1.84	&	1.64	$\pm$	0.17	&	0.07	$\pm$	0.05	&	0.94	\\
060906	&	3.68	&	43.6	&	2.44&	1.31	$\pm$	0.33	&	0.60	$\pm$	0.45	&	1.83	\\
061121	&	1.31	&	81	&	1.62	&	1.90	$\pm$	0.44	&	1.11	$\pm$	0.18	&	2.04	\\
070110	&	2.35	&	85	&	2.11	&	2.13	$\pm$	0.04	&	0.78	$\pm$	0.40	&	0.67	\\
070306	&	1.497	&	209.5	&	2.29	&1.63	$\pm$	0.39	&	0.28	$\pm$	0.21	&	1.02	\\
070318	&	0.836	&	74.6	&	1.4	&	0.10	$\pm$	0.04	&	0.24	$\pm$	0.58	&	0.16	\\
070721B	&	3.626	&	340	&	1.48	&	0.64	$\pm$	0.16	&	5.51	$\pm$	5.93	&	2.91	\\
071021	&	5	&	225	&	2.12	&	    1.81	$\pm$	0.43	&	0.60	$\pm$	1.08	&	1.66	\\
080310	&	2.4266	&	365	&	2.85	&	1.73	$\pm$	0.43	&	0.41	$\pm$	0.53	&	1.00	\\
080430	&	0.767	&	16.2	&	2.42	&0.71	$\pm$	0.17	&	0.05	$\pm$	0.02	&	0.06	\\
080607	&	3.036	&	79	&	1.68	&	0.08	$\pm$	0.02	&	27.49	$\pm$	15.53	&	14.85	\\
080707	&	1.23	&	27.1	&	1.81	&0.61	$\pm$	0.15	&	0.04	$\pm$	0.04	&	0.07	\\
080905B	&	2.374	&	128	&	1.49	&	0.50	$\pm$	0.12	&	7.05	$\pm$	72.07	&	0.75	\\
081007	&	0.5295	&	10	&	3	&   	1.20	$\pm$	0.28	&	0.02	$\pm$	0.01	&	0.02	\\
081008	&	1.9685	&	185.5	&	1.91	&1.06	$\pm$	0.26	&	0.64	$\pm$	0.37	&	1.32	\\
090529	&	2.625	&	$>$100	&	2.5	&	2.59	$\pm$	0.65	&	0.05	$\pm$	0.14	&	0.34	\\
090618	&	0.54	&	113.2	&	2.11	&0.66	$\pm$	0.16	&	0.48	$\pm$	0.07	&	2.79	\\
090927	&	1.37	&	2.2	&	1.64	&	1.15	$\pm$	0.33	&	0.03	$\pm$	0.07	&	0.03	\\
091029	&	2.752	&	39.2	&	2	&	1.19	$\pm$	0.29	&	0.56	$\pm$	0.36	&	1.27	\\
100302A	&	4.813	&	17.9	&	2.28&	5.17	$\pm$	1.23	&	0.23	$\pm$	0.26	&	0.38	\\
100418A	&	0.624	&	7	&	4.29	&	8.80	$\pm$	2.08	&	0.002	$\pm$	0.001	&	0.01	\\
100621A	&	0.542	&	63.6	&	2.15	&1.33	$\pm$	0.32	&	0.08	$\pm$	0.04	&	0.56	\\
100704A	&	3.6	&	197.5	&	2.6	&	    1.10	$\pm$	0.27	&	3.36	$\pm$	1.27	&	4.80	\\
100814A	&	1.44	&	174.5	&	1.9	&	14.45	$\pm$	3.38	&	0.26	$\pm$	0.08	&	1.60	\\
100906A	&	1.727	&	114.4	&	2.15	&0.69	$\pm$	0.17	&	1.33	$\pm$	0.72	&	2.94	\\
\bottomrule
\end{longtable}
From this table, we can find that the redshift of our
GRB sample is in the range of 0.42 (GRB 050803) to 5.11 (GRB
060522). And the mean value of redshift for these 43 bursts is about
2.3. The starting time ($t_1$) and the mid-point flux ($f_{\rm p}$)
of the plateau segment are obtained by the fitting of the
steep-to-shallow decay segment with a smoothed broken power law function
(Cui et al. 2010). The end time of this segment ($t_2$) is taken as
the break time between the plateau to normal decay phase. The
duration of plateau then is $\tau_{\rm GRB}=t_2-t_1$. With redshift
$z$, the luminosity distance ($D_{\rm L}$) of the burst can be obtained by adopting cosmological parameters
$\Omega_\mathrm{M}=0.3$, $\Omega_{\Lambda}=0.7$, and $H_0=71$ km
$\mathrm{s}^{-1}$ $\mathrm{Mpc}^{-1}$. Thus the luminosity in mid-point of plateau phase of GRB X-ray
afterglow then could be calculated by
\begin{equation}
L_{\rm plateau,GRB} = 4\pi\kappa_X \times D_{\rm L}^2\times f_{\rm p}.
\label{LGRB}
\end{equation}
Assuming the emission in the plateau phase from the source is mainly from the observed band ,the factor $\kappa_X$
corrects the flux at observational energy band ([$E_1, E_2$] in unit
of keV) of an instrument (XRT here, i.e., [$E_1$=0.3 keV, $E_2$=10
keV]) to that at a band $(0.01-100)/(1+z)$ keV which is
\begin{equation}
\kappa_X=\frac{\int^{100/(1+z)}_{0.01/(1+z)}E\Phi(E)dE}{\int^{E_2}_{E_1}E\Phi(E)dE},
\label{k}
\end{equation}
where $\Phi(E)\propto E^{-\Gamma_X}$, $\Gamma_X$ (as shown in Table \ref{tabGRB}) is the photon index for photon spectrum (Dainotti et al. 2010). And the error of $L_{\rm plateau,GRB}$ is deduced by the errors of the best fitting parameters for the plateau phase based on the error transfer formula.

As the opening angles for most of the GRB in our sample are not
known and the explosion of SN is thought to be almost
isotropic, here we take the gamma-ray isotropic energy of GRB as the
total energy of GRB in prompt explosion phase with observed fluence
$S$ and redshift $z$,
\begin{equation}
E_{\rm expl, GRB}=4\pi \kappa_{\gamma}
D_\mathrm{L}^2S/(1+z).\label{Eiso}
\label{Eexpl}
\end{equation}
 The factor $\kappa_{\gamma}$ is applied to convert the observed fluence at observational energy band of an instrument (from $E_1$ to $E_2$, in unit of keV) to that at a standard band $(1-10^4)/(1+z)$keV in rest frame of GRB (Bloom et al. 2001), which reads
\begin{equation}
\kappa_{\gamma}=\frac{\int^{10^4/(1+z)}_{1/(1+z)}EN(E)dE}{\int^{E_2}_{E_1}EN(E)dE},
\label{kgamma}
\end{equation}
where $E$ is photon energy, $N(E)$ is the band function defined by
Band et al. (1993). Since it's difficult to get the spectral index
for individual GRBs only from BAT observation with narrow energy
band, mean spectral indices $\alpha \simeq -1$, $\beta
\simeq -2.2$ and peak energy $E_{\rm p}\simeq 250$ keV obtained from the statistic are substituted into $N(E)$ formula (Preece et al.
2000).

For the type II SNe, three physical parameters: explosion energy
$E_{\rm expl, SN}$, envelope mass and initial radius are mainly
determined by the outburst properties: the plateau
duration $\tau$ in light curve, absolute V magnitude $M_V$ at
mid-plateau point, and the material velocity $u_{\rm ph}$ at the
photosphere. With these three observed parameters, Litvinova \&
Nadyozhin (1983, 1985, LN85 hereafter) have presented three
approximation formulae to calculate the three physical parameters
mentioned above based on the hydrodynamical models. We collect the
observed SNe II-P data with explosion energy $E_{\rm expl, SN}$ and
the bolometric luminosity in the mid-point of plateau phase $L_{\rm plateau,SN}$ . This
bolometric luminosity for our SNe
II-P sample comes from the work of Bersten \& Hamuy (2009, BH09
hereafter). They derived calibrations for bolometric corrections and
effective temperature from BVI photometry and obtained bolometric
light curve for a sample of 33 SNe II-P. Within this sample, only 11
SNe with the observed parameters has explosion energy $E_{\rm expl,
SN}$ deduced in the prior work (Nadyozhin 2003; Maguire et al.
2010). Our SNII-P sample are composed by these 11 SNe and their properties are presented in Table
\ref{tabSNe}.
\begin{table*}
 \centering
 \begin{minipage}{140mm}
  \caption{The properties of SNe II-P sample \label{tabSNe}}
  \begin{tabular}{@{}llrrrrlrlr@{}}
  \hline
  SN & c$z$ &$\tau_{\rm SN}$ & $L_{\rm plateau,SN}$ &$E_{\rm expl,SN}$  & References \\
& (km/s)&(day)& ($\times 10^{41}$erg/s)&($\times 10^{51}$ erg)&  \\
 \hline
1991al	&	4484	&	90	&	20.6	&	2.61	&	1, 7	\\
1992af	&	5438	&	90	&	12.5	&	2.46	&	1, 7	\\
1992ba	&	1165	&	100	&	7.5	&	0.57	&	1, 7	\\
1999br	&	1292	&	100	&	1.5	&	0.2	&	1, 7	\\
1999cr	&	6376	&	100	&	9.7	&	0.9	&	1, 7	\\
1999em	&	669	&	120	&	8	&	0.84	&	1, 8	\\
1999gi	&	592	&	115	&	6.7	&	0.64	&	 2, 8	\\
2003gd	&	657	&	113	&	7.8	&	1.04	&	3, 8	\\
2004dj	&	132	&	105	&	7.4	&	0.65	&	4, 8	\\
2004et	&	48	&	110	&	10.1	&	0.88	&	5, 8	\\
2005cs	&	463	&	118	&	3.1	&	0.17	&	 6, 8	\\
\hline \end{tabular}
 \begin{tablenotes}
  \item[*]References.¡ª(1) Hamuy (2001); (2) Nakano \& Kushida (1999); (3)Carnegie Type II Supernovae
Survey (CATS); (4) Vink¡äo et al. (2006); (5)Zwitter et al. (2004); (6) Kloehr et al. (2005); (7) Nadyozhin (2003); (8) Maguire et al. (2010)
\end{tablenotes}
\end{minipage}
\end{table*}
In BH09's sample, the zero point of time was taken as the middle point between
the plateau and the radioactive tail. We also take this zero point
as the end of the plateau phase in this work. A smoothed broken
power law is then used to fit the light curve for the data with
$t<0$:
\begin{equation}
L = L_0[(\frac{t}{t_{\rm p}})^{\omega \alpha_1}+(\frac{t}{t_{\rm
p}})^{\omega \alpha_2}]^{-1/\omega},
\end{equation}
where $L_0$ is the normalized parameter for the fitting. Parameter $\omega$ describes the sharpness of the break. $t_{\rm p}$ is
the time of beginning point of plateau. $\alpha_1$ and $\alpha_2$ present
the slopes of components before plateau and plateau. The duration of
plateau $\tau_{\rm SN}$ and the bolometric luminosity at the mid-point of
plateau $L_{\rm plateau,SN}$ can be obtained finally by the best
fittings parameters with $\tau_{\rm SN}=|t_{\rm p}|$ and $L_{\rm
plateau,SN}=L(t_{\rm p}/2)=L_0(2^{-\omega \alpha_1}+2^{-\omega
\alpha_2})^{-1/\omega}$.

\section{Results}
\label{sect:Results} The properties of GRB sample from the
observations and the parameters deduced from the formulae as described in
Equations (\ref{LGRB}) to (\ref{kgamma}) are presented in Table
\ref{tabGRB}: the redshift $z$, the duration of prompt phase
$T_{90}$, photon index in the afterglow phase $\Gamma_X$, the
duration of plateau $\tau_{\rm GRB}$, the bolometric luminosity at
the mid-point of plateau $L_{\rm plateau,GRB}$, and the explored energy
in the prompt phase $E_{\rm expl,GRB}$. Table \ref{tabSNe} lists the
properties of 11 SNe II-P included in this study. The selection standards for our SNe II-P sample are these SNe II-P with (1) the
measurement of the plateau duration (as shown in Column 3 of Table
\ref{tabSNe}); (2) the bolometric corrections for the lightcurves
(e.g. BH09), and then the bolometric luminosity $L_{\rm plateau,SN}$ at
the mid-point of plateau phase (as shown in Column 4 of Table
\ref{tabSNe}); (3) absolute V magnitude $M_V$
at plateau; (4) the material velocity $u_{\rm ph}$ at the
photosphere at mid-plateau point. Based on the parameters as
described by (1), (3), and (4), the explosion energy $E_{\rm
expl,SN}$ (as shown in Column 5 of Table
\ref{tabSNe}) can be obtained applying the hydrodynamical models as
presented by LN85.

\begin{figure}
  \centering
    \includegraphics[width=8cm]{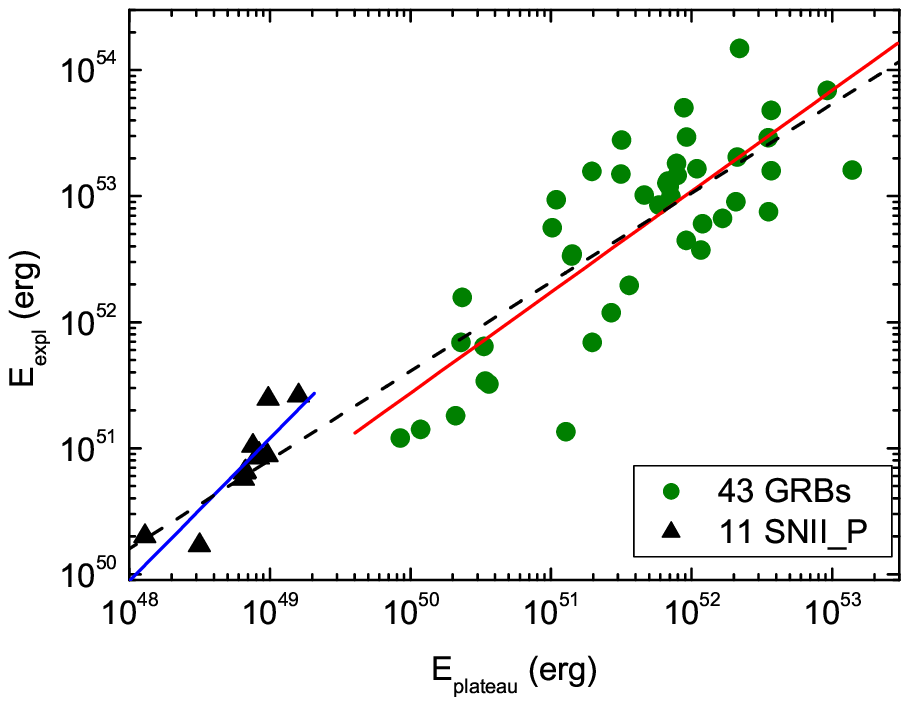}
    \includegraphics[width=8cm]{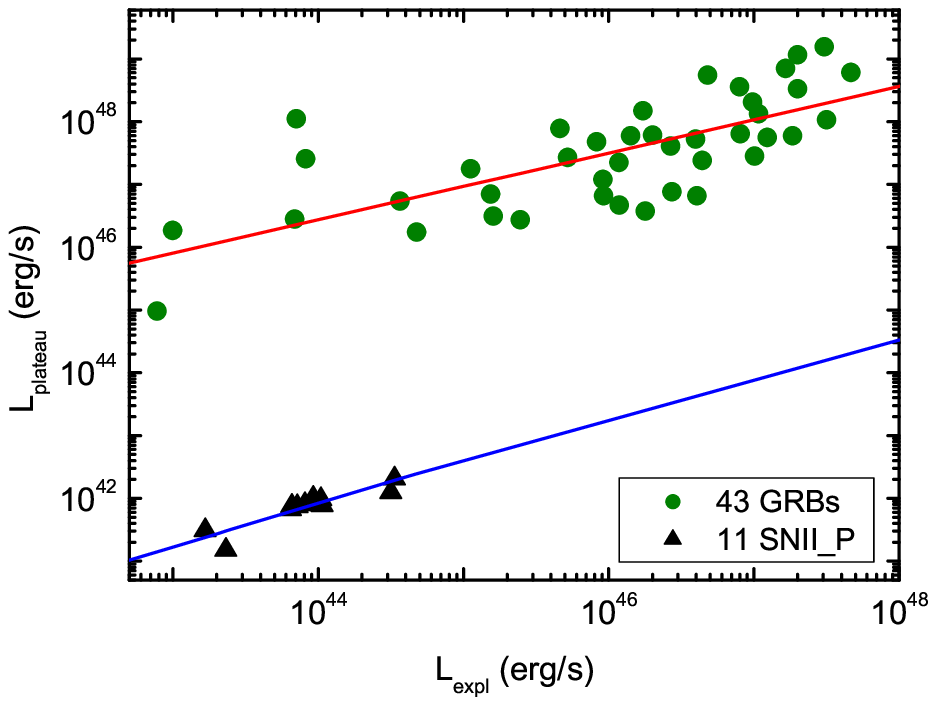}
    \caption{Correlation diagram of bolometric luminosities and energies at the mid-point of plateau phase and prompt phase for GRB and SNe II-P samples. The red and blue lines are the best linear fits for GRB and SNe II-P samples, respectively. Upper panel: The relation for energies $E_{\rm expl}$ in the prompt phase to that in the plateau phase $E_{\rm plateau}$, the dash line is the best fitting for both samples; Lower panel: The relation for the luminosities $L_{\rm plateau}$ to the ratio of $L_{\rm expl}$.
\label{Corfig}}
\end{figure}

Left panel of Figure \ref{Corfig} shows the correlation of product,
$\tau \times L_{\rm plateau}$, to the energy $E_{\rm expl}$ for GRB
X-ray afterglow and SNe II-P samples. Right panel of this figure
presents the relation of luminosities $L_{\rm plateau}$ and $L_{\rm
expl}$ (i.e. $E_{\rm
expl}/\tau$). Linear fit is applied to test the correlations for each
sample in the logarithm coordinate and fitting results are presented
in Table \ref{tabCor}. From this table and Figure \ref{Corfig}, we
can find that $E_{\rm plateau}$ (i.e. $\tau \times L_{\rm plateau}$) and $E_{\rm expl}$ as well
as $L_{\rm plateau}$ and $L_{\rm expl}$ are correlated for two samples, respectively. All the Spearman correlation
coefficients $r$ are larger than 0.8 with chance probabilities
$p\sim10^{-4}$. This implies that the prompt isotropic gamma-ray
energy is indeed correlated with the isotropic X-ray energy in the
plateau phase (Liang et al. 2007). And the energy budgets for the
plateau phase and the (prompt) explosion energy are correlated for
both samples. The slopes in the $E_{\rm expl}$--$E_{\rm
plateau}$ diagram n the logarithm coordinate for two linear fittings are $0.80\pm0.09$ and
$1.13\pm0.20$. For the sample (SNe II-P+GRB), it's $0.73\pm0.14$.
Thus we can find that all of slopes, i.e. the power indies in linear coordinate, are very near. And the slopes of the correlation $L_{\rm
plateau}$--$L_{\rm expl}$ in the logarithm coordinate are also found to be very closer ($0.79\pm0.07$ and
$0.69\pm0.11$ for GRB and SN II-P samples, respectively. Thus it's
possible that the processes of energy injected to the shock/ejected
material in the (prompt) explosion and the plateau phase are very
similar.
.
\begin{table*}
 \centering
 \begin{minipage}{140mm}
  \caption{Fitting results for GRB and SNe II-P samples.\label{tabCor}}
  \begin{tabular}{@{}llrrrrlrlr@{}}
  \hline
 correlation &Sample & Slope & $r^a$& $SD^b$& $p^c$\\
 \hline
\multirow{3}{*}{$E_{\rm expl}$--$E_{\rm plateau}$} &GRB &0.80 (0.09) &0.81 &0.47  &$<10^{-4}$\\
\cline{2-6}
&SNe II-P & 1.13 (0.20) &0.89 &0.18 &2.93$\times 10^{-4}$ \\
\cline{2-6}
&SNe II-P + GRB & 0.73 (0.14) &0.91 &0.43 &$< 10^{-4}$ \\
\hline
\multirow{2}{*}{$L_{\rm plateau}$--$L_{\rm expl}$} &GRB &0.79 (0.07) &0.88 &0.44  &$<10^{-4}$\\
\cline{2-6}
&SNe II-P & 0.69 (0.11) &0.91 &0.14 &1.26$\times 10^{-4}$ \\
\hline \end{tabular}
 \begin{tablenotes}
  \item[*]$a$: Spearman correlation coefficient
  \item[*]$b$: Standard deviation
  \item[*]$c$: Chance probability
\end{tablenotes}
\end{minipage}
\end{table*}

The gap ($\sim 2.23$) in vertical direction between the best
fitting lines of two sample in $L_{\rm plateau}$--$L_{\rm expl}$
diagram might indicate that the different ways or levels to provide
energy for explosions during X-ray plateau for GRB afterglow and
(prompt) plateau for SNe II-P. That is to say, the central engine activities
or energy budgets of GRB and SNe II-P during the plateaus could be
different. The energy poured into the ejecta or circum-burst
materials for GRB is larger than that for SNe II-P during the
plateau phases. Therefore, the plateaus for two samples would
manifest in different observational energy bands. For GRB, it
exhibits in X-ray band but for SNe II-P in optical band.  The very
near power law indies of best fittings for GRB and SNe II-P samples
considering the error bars may indicate that the hydrodynamic process or the energy injection behavior,
e.g. the shockwave propagation in the circum-materials around the
burst, during the plateau phase for GRB and SNe II-P could be very
similar.

\section{Conclusion and discussions}

With a comparative work about the plateau in the lightcurves of GRB
X-ray afterglow and in the explosion phase of SNe II-P, we find that the (prompt) explosion
energy $E_{\rm expl}$ and the product between the mid-point
bolometric luminosity in the plateau phase $L_{\rm plateau}$ and the
duration of plateau $\tau$, as well as the luminosity $L_{\rm plateau}$
and $L_{\rm expl}$ for two samples are correlated,
respectively. All the Spearman correlation coefficients for the
linear fittings in $E_{\rm expl}$--$E_{\rm plateau}$ and
$L_{\rm plateau}$--$E_{\rm expl}/\tau$ diagrams with logarithm
coordinate are larger than 0.8 with chance probabilities
$p\sim10^{-4}$. This implies that the energy injected in the
(prompt) explosion and plateau phases are correlated for GRB and
type II-P SNe, respectively. The similar power indies of the best
fittings for two samples may indicate a
similar hydrodynamics during the energy injection in the
plateau phases. The gap in $L_{\rm plateau}$--$L_{\rm
expl}$ diagram between two best fitting lines might imply
that the center engine or the style to poured energy into ejecta of
two samples could be different.

The optical data of GRB afterglow have been collected by Li et al. (2012). An optical shallow-decay segment in these GRB afterglow is observed in 39 GRBs. Based on their results, a rough proportionality between the isotropic energy in the prompt phase $E_{\gamma,iso}$ and isotropic R-band energy $E_{R, iso}$ in the optical shallow-decay segment is observed in their work. And the best fitting between these two quantities is $\log E_{R, iso}=0.40+0.47 \log E_{\gamma,iso}$ with chance probability $p\sim 6 \times 10^{-3}$. The isotropic energy $E_{\gamma,iso}$ is the same as the energy $E_{\rm expl}$ in prompt phase as presented in equation (\ref{Eexpl}) in this work. Compared with the fitting slopes shown in Table (\ref{tabCor}), we can find that the correlation between $E_{R, iso}$ and $E_{\gamma,iso}$ is different from that of $E_{\rm plateau}$ and $E_{\rm expl}$ in this work.

The very origin of plateau is quite difficult to identify though it is very likely related to the external shock (e.g. Zhang
2007). However, the spectral index generally does not change across
the temporal break (Liang et al. 2007) from the plateau phase to the
following decay phase. Thus the models invoking
radiation mechanism can be ruled out for the origin of plateau
phase. A hydrodynamical or geometrical origin is proposed by Zhang
(2007). The continuous injection dynamics was discussed invoking a
spin-down pulsar (Dai \& Lu 1998a, b; Zhang \& M{\'e}sz{\'a}ros
2001) with a smoothly varying luminosity $L \propto t^{-q}$ (Zhang
\& M{\'e}sz{\'a}ros 2001) and a value $q\sim 2$ is suggested by the
observational data (Fan \& Xu 2006; Rowlinson et al. 2010).
Alternatively, the GRB plateau may be due to the solidification of
quark stars (Xu \& Liang 2009; Dai et al. 2011), that favor clean
fireballs without baryon contamination (Paczynski \& Haensel 2005;
Cheng et al. 2007).

The hydrodynamical process of envelope ejection is also one of the
characteristic features of SNe II-P. Litvinova \& Nadezhin (1983,
1985) presented a series of hydrodynamical models of SNe II-P and
found that the light curves was determined by the size and mass of
unstable progenitor envelope. The usual hypothesis about the SNe
explosion can be decoupled into the collapse of the core and the
ejection of the envelope (e.g. Grassberg et al. 1971, Woosley 1988).
These two parts are independent and the observations are only
determined by the propagation process of the shock wave producing
from the core collapse through the envelope (Falk \& Arnett 1977;
Bersten et al. 2011). If the process of shock wave propagation in
the envelope is the same as that of external shock involving GRB
afterglow plateau, the hydrodynamics of the energy injection about the plateau phenomena may
be similar for GRB afterglow and SNe II-P samples. And the timescale
of the similarly and underlying hydrodynamical process and energy levels
poured from the center object could be different because they are
possibly determined by the time and budget of the energy injection
from center engine. And thus it's possible that the plateaus for GRB
afterglow and for SNe II-P exhibit in different energy band.

\begin{acknowledgements}
We would like to thank useful discussions at our pulsar group of PKU. This work is supported by the National Basic Research Program of China
(Grant Nos. 2012CB821800, 2009CB824800), the National Natural Science Foundation of China (Grant Nos. 11225314, 11103026, 10935001).
\end{acknowledgements}

\label{lastpage}


\begin{thebibliography}{99}
%

  \bibitem[Arnett(1980)]{arne80} Arnett, W. D. 1980, ApJ, 237, 541
\bibitem[Band et al.(1993)]{band93}Band, D. L., Matteson, J., Ford, L., et al. 1993, ApJ, 413, 281
\bibitem[Barbon et al.(1979)]{barb79} Barbon, R., Ciatti, F., \& Rosino, L. 1979, A\&A, 72, 287
 \bibitem[Bersten \& Hamuy(2009)]{bers09} Bersten , M. C. \& Hamuy, M., 2009, ApJ, 701, 200
 \bibitem[Bersten et al.(2011)]{bers11} Bersten, M. C., Benvenuto, O., \& Hamuy, M. 2011, ApJ, 729, 61
\bibitem[Bloom et al.(2001)]{bloom01} Bloom, J. S., Frail, D. A., \& Sari, R. 2001, AJ, 121, 2879
 \bibitem[Burrows et al.(2003)]{burr03} Burrows, D. N., Hill, J. E., Nousek, J. A. et al. 2003, Proc. SPIE, 4851, 1320
\bibitem[Burrows et al.(2006)]{burr06} Burrows, A., Livne, E., Dessart, L., Ott, C. D., \& Murphy, J. 2006, New Astron.
Rev., 50, 487
\bibitem[Chen et al.(2007)]{chen07} Chen, A. B., Yu, T. H., \& Xu, R. X. ApJ, 2007, 668, L55
\bibitem[Cui et al.(2010)]{cui10} Cui, X. H., Liang, E. W., Lv, H. J., Zhang, B. Bi., \& Xu, R. X., 2010, MNRAS, 401, 1465
\bibitem[de Pasquale et al.(2009)]{de09} de Pasquale, M., Evans, P., Oates, S., et al. 2009, MNRAS, 392, 153
\bibitem[Dai et al.(2011)]{dai11} Dai, S., Li, L. X., \& Xu, R. X., 2011, Science in China Series G: Physics, Mechanics \& Astronomy,  accepted(arXiv:1008.2568v1)
\bibitem[Dai \& Lu(1998a)]{dai98a} Dai, Z. G., \& Lu, T., 1998a, PRL, 81, 4301
\bibitem[Dai \& Lu(1998b)]{dai98b} Dai, Z. G., \& Lu, T., 1998b, A\&A, 333, L87
\bibitem[Dainotti et al.(2010)]{dai10} Dainotti, M. G., et al. 2010, ApJ, 722, L215
\bibitem[Falk \& Arnett(1977)]{falk77} Falk, S. W. \& Arnett, W. D. 1977, ApJS, 33, 515
\bibitem[Fan \& Piran(2006)]{fanP06} Fan, Y. Z. \& Piran, T.  2006, MNRAS, 369, 197
\bibitem[Fan \& Xu(2006)]{fan06} Fan, Y. Z. \& Xu, D., 2006, MNRAS, 372, L19
\bibitem[Genet et al.(2007)]{genet07} Genet, F., Daigne, F., \& Mochkovitch, R., 2007, MNRAS, 381, 732
\bibitem[Granot et al.(2006)]{granot06} Granot, J., K\"{o}nigl, A., \& Piran, T., 2006, MNRAS, 370, 1946
\bibitem[Grassberg et al.(1971)]{gras71} Grassberg, E. K., Imshennik, V. S., \& Nadyozhin, D. K. 1971, Ap\&SS, 10, 28
\bibitem[Hamuy(2001)]{Hamuy01} Hamuy, M. 2001, PhD thesis, Univ. Arizona
\bibitem[]{} Hjorth, J. \& Bloom, J., Cambridge Astrophysics Series, 2012, 51, 169
\bibitem[Janka et al.(2007)]{Janka07} Janka, H. T., Langanke, K., Marek, A., et al.
2007, Phys. Rep., 442, 38
\bibitem[]{} Klu¡äzniak, W., \& Ruderman, M. 1998, ApJ, 505, L113
\bibitem[]{} Li, L., Liang, E., Tang, Q., et al. 2012, ApJ, 758, 27
\bibitem[]{} Li, L. X. 2006, MNRAS, 372, 1357
\bibitem[]{} Li, L. X. 2008, AIPC, 1065, 273
\bibitem[Liang et al.(2007)]{liang07} Liang, E. W., Zhang, B. B., \& Zhang, B.  2007, ApJ, 670, 565
\bibitem[Litvinova \& Nadezhin(1983)]{litv83} Litvinova, I. I. \& Nadezhin, D. K. 1983, Ap\&SS, 89, 89
\bibitem[Litvinova \& Nadezhin(1985)]{litv85} Litvinova, I. Y. \& Nadezhin, D. K. 1985, Soviet Astronomy Letters, 11, 145
\bibitem[Maguire(2010)]{magu10} Maguire, K., di Carlo, E., Smartt, S. J. et al. 2010, MNRAS, 404, 981
\bibitem[Nadyozhin(2003)]{nady03} Nadyozhin, D. K., 2003, MNRAS, 346, 97
\bibitem[Nousek et al.(2006)]{nous06} Nousek, J. A., Kouveliotou, C., Grupe, D., et al. 2006, ApJ, 642, 389
\bibitem[Paczynski \& Haensel(2005)]{pacz05} Paczynski, B. \& Haensel, P. MNRAS, 2005, 362, L4
\bibitem[Panaitescu et al.(2006)]{pana06} Panaitescu, A., M{\'e}sz{\'a}ros, P., Burrows, D., et al. 2006, MNRAS, 369, 2059
\bibitem[Popov(1993)]{Popov93} Popov, D. V. 1993, ApJ, 414, 712
\bibitem[Preece et al. (2000)]{Pree00} Preece, R. D., Briggs, M. S., Mallozzi, R. S. et al. 2000, ApJS, 126, 19
\bibitem[Pskovskii(1978)]{Pskovskii78} Pskovskii, Iu. P. 1978, SvA, 22, 201
\bibitem[Rees \& M{\'e}sz{\'a}ro(1998)]{Rees98} Rees, M. J. \& M{\'e}sz{\'a}ros, P., 1998, ApJ, 496, L1
\bibitem[Rowlinson et al.(2010)]{rowl10} Rowlinson, A., O'Brien, P. T., Tanvir, N. R., Zhang, B. et al. 2010, MNRASs, 409, 531
\bibitem[]{} Soderberg, A. M. et al., 2004, ApJ, 606, 994
\bibitem[]{} Soderberg, A. M. et al., 2006, Nature, 442, 1014
\bibitem[]{} Soderberg, A. M. et al., 2008, Nature, 453, 469
\bibitem[Shao \& Dai(2005)]{shao05} Shao, L. \& Dai, Z. G. 2005, ApJ, 633, 1027
\bibitem[Shao \& Dai(2007)]{shao07} Shao, L. \& Dai, Z. G. 2007, ApJ, 660, 1319
\bibitem[Uhm \& Beloborodov(2007)]{uhm07} Uhm, Z. L. \& Beloborodov, A. M. 2007, ApJ, 665, L93
\bibitem[Utrobin \& Chugai(2009)]{utro09} Utrobin, V. P. \& Chugai, N. N. 2009, A\&A, 506, 829
\bibitem[Woosley(1988)]{woos88} Woosley, S. E. 1988, ApJ, 330, 218
\bibitem[]{} Woosley, S. E. 1993, ApJ, 405, 273
\bibitem[]{} Woosley, S. E., \& Bloom, J. S. 2006, ARA\&A, 44, 507
\bibitem[Xu \& Huang(2011)]{xu11} Xu, M. \& Huang, Y. F. 2011, arXiv:1103.3978
\bibitem[Xu \& Liang(2009)]{xu09} Xu, R. X. \& Liang, E. W., 2009, Science in China Series G: Physics, Mechanics \& Astronomy, 52, 315
\bibitem[Yamazaki(2009)]{Yamazaki09} Yamazaki, R., 2009, ApJ, 690, 118
\bibitem[Young \& Branch(1989)]{Young89} Young, T. R. \& Branch, D. 1989, ApJ, 342, L79
\bibitem[Zhang \& M{\'e}sz{\'a}ros(2001)]{zhang01} Zhang, B. \& M{\'e}sz{\'a}ros, P. 2001, ApJ, 552, L35
\bibitem[Zhang et al.(2006)]{zhang06} Zhang, B., Fan, Y. Z., Dyks, J., et al. 2006, ApJ, 642, 354
\bibitem[Zhang(2007)]{zhang07} Zhang, B. 2007, Chin. J. Astron. Astrophys., 7, 1

\end{thebibliography}
\end{document}